# From Offline to Inline Without Pain

## A Practical Framework for Translating Offline MR Reconstructions to Inline Deployment Using the Gadgetron Platform

## Table of Contents

# Title page

**Title:**

From Offline to Inline Without Pain: A Practical Framework for Translating Offline MR Reconstructions to Inline Deployment Using the Gadgetron Platform

**Authors:**

Zihan Ning[1], Yannick Brackenier[1], Sarah McElroy[1,2], Sara Neves Silva[1], Lucilio Cordero-Grande[3], Sam Rot[4], Liane S. Canas[1], Rebecca E Thornley[1,5], David Leitão[1], Davide Poccecai[1], Andrew Cantell[1], Rene Kerosi[1], Anthony N Price[1,5], Jon Cleary[1,5], Donald J Tournier[1], Jana Hutter[1,5], Philippa Bridgen[5,6], Pierluigi Di Cio[5,6], Michela Cleri[5,6], Inka Granlund[5,6], Lucy Billimoria[5,6], Yasmin Blunck[7], Shaihan Malik[1],Marc Modat[1], Claire J Steves[5,8], Joseph V Hajnal[1]

**Affiliations:**

[1]Imaging physics and engineering research department, School of Biomedical Engineering and Imaging Sciences, Kings College London, London, United Kingdom

[2]MR Research Collaborations, Siemens Healthcare Limited, Camberley, UK

[3]Biomedical Image Technologies, ETSI Telecomunicación, Universidad Politécnica de Madrid & CIBER-BBN, ISCIII, Madrid, Spain

[4]NMR Unit, Queen Square MS Centre, Queen Square Institute of Neurology, University College London, London, United Kingdom

[5]Guy's and St Thomas' NHS Foundation Trust, London, United Kingdom

[6]London Collaborative Ultra high field System (LoCUS), Kings College London, London, United Kingdom

[7]Department of Biomedical Engineering, The University of Melbourne, Parkville, Australia

[8]Department of Twin Research and Genetic Epidemiology, King's College London, London, United Kingdom

**Correspondence:**

Correspondence to Zihan Ning, PhD:

Address: Department of Perinatal Imaging and Health, Floor 1, South Wing, St Thomas' Hospital, Westminster Bridge Road, London, SE1 7EH, United Kingdom

Tel: +44-07436789013

Email: zihan.1.ning@kcl.ac.uk

# Abstract

**Purpose:**

To develop and validate a practical, open-source framework to overcome common issues in inline deployment of established offline MR reconstruction, including (1) scan disruption or delay from lengthy reconstructions, (2) limited support for multi-scan input reconstructions, (3) the need to adapt scripts for different raw data formats, and (4) limited guidance and experience in retaining scanner reconstructions and applying scanner-based post-processing to custom-reconstructed images.

**Methods:**

The framework builds upon the Gadgetron platform and includes: (1) a general input converter to transform Gadgetron-used ISMRMRD format raw into a Siemens format raw structure, facilitating reuse of existing code; (2) an asynchronous trigger-and-retrieve mechanism enabling long custom reconstructions without delaying scanner processes; (3) resource-aware scheduling for parallel execution of reconstructions; (4) integrated file management to support multi-scan inputs; and (5) preservation of scanner-based reconstructions and post-processing. The framework was validated on 2 Siemens scanners for SENSE, AlignedSENSE, and NUFFT reconstructions, and in a large-cohort study.

**Results:**

Minimum code modification for inline deployment has been demonstrated, and all reconstructions were successfully executed inline without disrupting scanner workflows. Images were retrieved via automated or retro-reconstruction, with scanner-based post-processing applied to custom outputs. Multi-sequence reconstructions were executed using GPU-aware scheduling, confirming feasibility for routine and large-scale applications. In 480 consecutive examinations, inline reconstructions were retrieved in 99% of cases without disruptions.

**Conclusion:**

The framework lowers the technical barrier to inline deployment of offline reconstructions, enabling robust, scalable, and post-processing-compatible integration. It is openly available with documentation and demonstration cases to support reproducibility and community adoption.

# Introduction

Advanced MR reconstruction and post-processing techniques are progressing rapidly, but adoption in clinical settings and large-scale research remains limited due to challenges in integration with normal scanner workflows. Without this inline integration, executing custom algorithms demands considerable effort. Extraction of raw (k-space) data and transmission to a suitable compute resource is generally manual, return of outputs to the scanner database is often infeasible and import to a suitable PACS (Picture Archiving and Communication System) can be cumbersome. Offline processes may not allow for standard post-processing (e.g., bias field and distortion correction), limiting compatibility with other acquired images.

Hitherto, integration of custom reconstructions into seamless workflows has required researchers to work within proprietary native scanner frameworks. Although feasible under suitable research agreements, it has required specialist knowledge and often considerable reworking of code. There are open-source platforms, such as Yarra[1] and Gadgetron[2], designed for prototyping and integration of reconstructions into workflows, and these have the advantage of supporting widely used high-level languages like Python and MATLAB, reducing reimplementation effort. Yarra[1] is a semi-inline solution that automates raw data collection, custom reconstruction triggering, and PACS or workstations export, though it lacks inline image review capabilities. In contrast, Gadgetron supports inline custom reconstructions including returning images directly to the scanner console[2,3]. It also functions as a toolbox with modular components for reconstruction and post-processing. More recently, vendors have begun embracing open-source toolboxes (e.g., Gadgetron and BART) by offering hybrid solutions integrated into their proprietary workflows, such as FIRE™ [5] and OpenRecon™ [6] for Siemens, GyroTools® for Philips[7], and Orchestra Live™ for GE[8].

Whereas DICOM provides a universal interchange format for medical images, there has historically been no equivalent for k-space data. ISMRMRD format[9] provides a step towards open standards, is native for Gadgetron and has been adopted by some vendors for their next generation frameworks (e.g. Siemens[10] FIRE™ and OpenRecon™, GE[11] Orchestra™, and Philips[12] Recon 2.0™). However, currently most existing algorithms have been written to accept only proprietary data formats.

There is a growing range of available options for creating inline reconstruction workflows, but many practical challenges remain. Common issues include: (1) scanning disruption or delay if inline reconstructions are time-consuming; (2) limited support for reconstructions requiring multiple inputs from different scans; (3) the need to adopt different raw data formats; (4) limited guidance and experience in retaining scanner reconstructions for early review or comparison, and in applying scanner-based post-processing to custom-reconstructed images.

To address these challenges, we developed a practical open-source framework based on Gadgetron. The aim is to enable inline deployment of established offline MR reconstructions, regardless of computation time or the number and source of inputs, while requiring minimal adaptation to the original scripts. Full inline operation places custom-reconstructed images directly into the native scanner database, which is ideal for integration with established workflows and standard data management pathways, such as connection to PACS. There are governance requirements associated with such complete integration, and it is critical to acknowledge and address these appropriately, but how that is done is beyond the scope of this paper.

The proposed framework, which is initially implemented for Siemens scanners, incorporates the following key features:

- ***Maximum Script Reuse for Rapid Prototyping:*** Allows unchanged data input by converting ISMRMRD-formatted data into a Siemens Twix-like structure.
- ***Parallel and Non-Disruptive Execution:*** Supports an asynchronous trigger-and-retrieval mechanism[13] to avoid delays in subsequent scans and reconstructions, and uses multi-GPU scheduling with resource monitoring to run custom reconstructions efficiently in parallel.
- ***Integrated Multi-input Data Management***: Supports structured scan-origin tracking for all data files to enable input readiness checks for multi-input reconstructions, and ensuring reliable automatic image retrieval for all reconstructions.
- ***Preservation of Scanner Reconstruction and Post-processing:*** Enables early image review through retained scanner reconstructions, and supports scanner-based post-processing on custom outputs to ensure consistency.
- ***Robust Operation*****:** Real-time logging and error tracking ensure resilience and recovery without requiring rescans or interrupting acquisitions.

This framework lowers technical barriers for translating offline methods into inline workflows, enabling efficient and scalable deployment in clinical protocols and large-cohort studies. Validation was performed on three aspects: (1) feasibility for inline implementation of established offline reconstructions, (2) feasibility for multi-sequence implementation within one examination, and (3) robustness.

# Methods

## Physical Infrastructure and Framework Architecture

The proposed framework operates on a physical infrastructure comprising four core components (Fig. 1A): (1) the MR scanner (magnet, gradient and RF systems), (2) the MR console, (3) the MR reconstruction system with the Gadgetron program installed enabling related modules, and (4) an external reconstruction server with the Gadgetron client[3].

### *Standard Gadgetron pipeline*

Gadgetron[2] integrates with the native scanner reconstruction pipeline using an "emitter" module to send raw data to the external server and an "injector" module to return images. After acquisition, the emitter converts vendor-specific raw data into ISMRMRD format[9,14], using predefined markup documents called ParameterMaps [10], and transmits it to the server via TCP/IP connection. Intermediate scanner reconstruction modules are bypassed, thus no scanner-reconstructed images are generated. Instead, on the external server, a customized chain of Gadgetron programs, named "Gadgets", performs pre-processing and user-defined reconstruction, with support for Python and MATLAB implementations. After processing, final images are returned to the scanner reconstruction workflow at the injector. The images are stored in the scanner database, and made available for console review or export (e.g., to PACS). Since the scanner workflow waits for image return and operates sequentially, long reconstructions can delay subsequent tasks. Moreover, because workflows for different scans do not share memory, reconstructions requiring input from multiple scans are not supported.

### *Rapid Prototyping with the Framework*

Most reconstruction methods are first developed offline using vendor-specific raw data, which are generally incompatible with ISMRMRD during inline translation. On Siemens platforms, the ISMRMRD output from the Gadgetron emitter differs significantly from the original "Twix" format in both header structure and k-space organization. Furthermore, only limited headers are converted by default, and standard preprocessing by default Gadgets (e.g., denoising, gridding) are required for data collection but may alter the data or consume dependent lines (e.g., noise scans), further increasing the mismatch. Consequently, existing reconstruction scripts often require extensive modification for inline deployment.

To address these issues, the proposed framework omits default Gadgetron preprocessing steps, deploying only the Accumulate Gadget to collect raw k-space lines (Fig. 1B). An updated ParameterMap ensures that key Twix headers are preserved in the ISMRMRD output. A general input converter is then provided and applied to reconstruct a Twix-like structure. It restores the expected header structure, populates key fields and reorganizes the k-space lines into a matrix format consistent with outputs from Siemens' official reader (i.e., mapVBVD).

### *Multi-input processing and asynchronous operation*

To support long and/or multi-input reconstructions, the framework employs an asynchronous trigger-and-retrieval mechanism[13] with parallel processing and integrated data management (Fig. 1B, C). Each reconstruction involves a "target scan" to acquire data and trigger reconstruction, and a "retrieval scan" to return images back to the console. Alternatively, Siemens' retro-reconstruction feature can be used for image retrieval after the examination session.

During the target scan, k-space data are streamed to the external server via the emitter module. A Read&Save handler receives the ISMRMRD-formatted streaming data, applies the general input converter to reconstruct a Twix-like structure, and saves it to an organized folder named with subject ID and date. Files are tagged with category (raw data or image to retrieve), sequence name, and timestamp to prevent overwriting. After saving, the handler exits and triggers a control script, while the Gadgetron workflow completes without returning images to the injector, allowing the scanner reconstruction pipeline to proceed without delay. If immediate image feedback is required and native reconstruction is feasible, the emitter can retain intermediate modules, allowing parallel scanner-based reconstruction.

The control script then checks two conditions before launching the reconstruction: (1) availability of all required inputs and (2) sufficient computational resources (e.g., GPU). If both are met, the custom reconstruction is launched; otherwise, the process is automatically queued until conditions are satisfied. Two log files are maintained: a resource log for tracking resource usage, and a computation log for monitoring progress and errors.

For image retrieval, either a retrieval scan (i.e., a short dummy scan (<5 seconds)) or retro-reconstruction can be used. In both cases, the emitter sends a matrix-size-aligned raw dataset to the external server but discarded; instead, saved images are located, loaded, and returned via the injector to the MR reconstruction workflow (Fig. 1C). The injector precedes scanner post-processing modules to ensure consistency with native images. To compare the two methods, retrieval scans automate image return within the protocol but require dedicated settings (e.g., matching matrix size and planning position), whereas retro-reconstruction inherently preserves these settings but requires manual pipeline selection on the MR console interface (see Supplementary Material A for detailed comparisons).

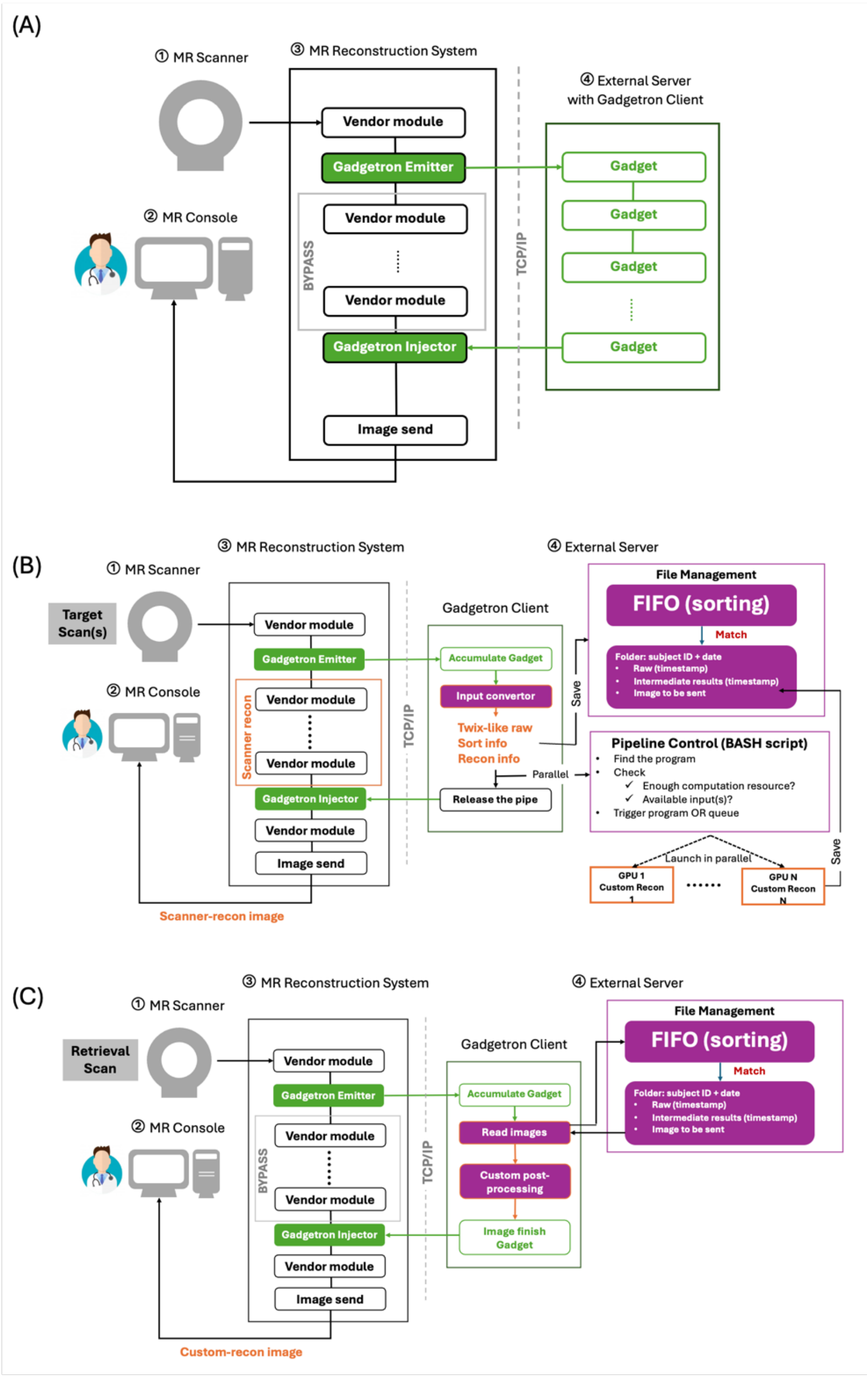

*Figure 1 The diagram illustrates the conventional Gadgetron workflow (A) and the workflow with the proposed framework, including the target scan (B) and retrieval scan/retro-reconstruction (C). The proposed framework was built upon four core components: the MR scanner, the MR console, the MR reconstruction system with the Gadgetron program (enabling the emitter and injector modules), and an external reconstruction server with the Gadgetron client. For the target scan, the framework triggers a parallel pathway for raw data conversion, saving, and reconstruction, while preserving the conventional scanner reconstruction for early review on the scanner console. Once custom reconstructions complete on*

*the external server, images can be retrieved to the console via retrieval scans or retro-reconstructions, with scanner post-processing applied.*

## Framework Validation

### *Feasibility for inline implementation of established offline reconstructions*

Three pre-existing reconstruction algorithms have been implemented inline via the proposed framework: (1) conventional SENSE with linear Cartesian trajectory[15]; (2) AlignedSENSE[16] with the DISORDER trajectory[17] for motion correction reconstruction using a rigid body motion model; and (3) NUFFT[18] with radial trajectory data for $^{23}Na$ imaging[19,20]. For SENSE and AlignedSENSE reconstruction, coil sensitivity estimation was performed by ESPIRiT[21] with options for embedded auto-calibration lines or external reference scans, with the latter used to demonstrate multi-input reconstructions. The detailed information of inline deployments about scanner, external servers and sequence parameters are provided in Supplementary Material Table S1 and S2.

### *Feasibility of multiple inline-reconstructions within one examination*

To evaluate the framework's capabilities to achieve multiple asynchronous inline reconstructions within a single examination, a healthy volunteer was imaged with the protocol: (1) an external reference scan for ESPIRiT-based coil sensitivity estimation[21]; (2) T1-MPRAGE, (3) FLAIR, (4) SWI, and (5) T2W SPACE, all with inline implemented AlignedSENSE reconstruction (detailed parameters in Supplementary Table S3). FLAIR and T2W SPACE used coil sensitivity maps from the external reference, while the others relied on embedded auto-calibration lines.

### *Robustness*

Framework robustness was validated in a large-scale cohort study, the TwinsUK Project[22,23], incorporating a motion-corrected T1-MPRAGE sequence with DISORDER trajectory and inline AlignedSENSE reconstruction. All scans acquired between 3 April 2024 and 4 April 2025 were reviewed to assess: (1) any disruption to scans or reconstructions, and (2) successful retrieval of both scanner- and AlignedSENSE-reconstructed images.

# Results

## Feasibility for Inline Implementation of Established Offline Reconstruction Scripts

The proposed framework is open-sourced along with the three demonstration cases, including both the original offline code and their inline versions, accompanied by a detailed user manual: https://github.com/ZihanNing/Practical_Inline_Recon_Framework-public.git.

To illustrate the ease of translation, Fig. 2A outlines the 4-step process for adapting a Twix-input offline reconstruction for inline use: (1) registering a new pipeline in the centralized configuration file, defining (2) Read&Save and (3) background reconstruction handlers from the templates, and (4) wrapping and adapting the original reconstruction script to be inserted in the background reconstruction handler. Since the ISMRMRD format raw has been converted back to a Twix-like structure, the original script could be largely reused. For minimal changes, replace vendor-specific data reading functions with access to the provided Twix-like input. For the three demonstrations, minimal modifications (% of code lines changed) required for inline implementation were: 0.07% (18/26693) for NUFFT, 0.19% (65/33972) for SENSE, and 0.04% (47/106522) for AlignedSENSE. Exact code-level differences are available in the repository[14]. In the open-sourced SENSE and AlignedSENSE versions, additional changes were made to support external reference scans (i.e., 0.84% (284/33972 ) for SENSE, and 0.40% (428/106522) for AlignedSENSE).

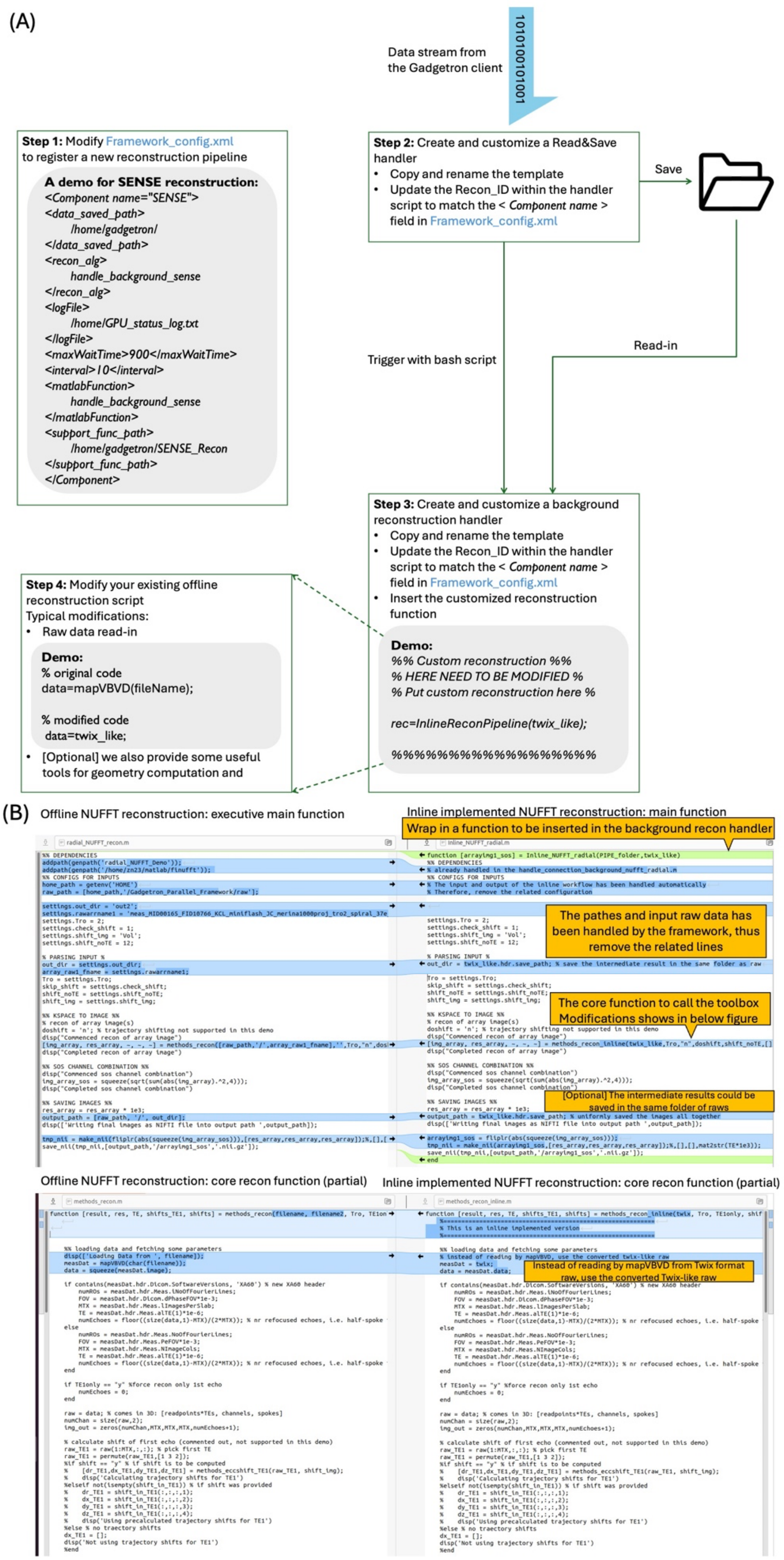
(A)
Data stream from the Gadgetron client
101010010100 1
Step 1: Modify Framework_config.xml to register a new reconstruction pipeline
A demo for SENSE reconstruction:
<Component name="SENSE">
<data_saved_path>
/home/gadgetron/
</data_saved_path>
<recon_alg>
handle_background_sense
</recon_alg>
<logFile>
/home/GPU_status_log.txt
</logFile>
<maxWaitTime>900</maxWaitTime>
<interval>10</interval>
<matlabFunction>
handle_background_sense
</matlabFunction>
<support_func_path>
/home/gadgetron/SENSE_Recon
</support_func_path>
</Component>
Step 2: Create and customize a Read&Save handler
• Copy and rename the template
• Update the Recon_ID within the handler script to match the < Component name > field in Framework_config.xml
Save
Trigger with bash script
Read-in
Step 3: Create and customize a background reconstruction handler
• Copy and rename the template
• Update the Recon_ID within the handler script to match the < Component name > field in Framework_config.xml
• Insert the customized reconstruction function
Demo:
%% Custom reconstruction %%
% HERE NEED TO BE MODIFIED %
% Put custom reconstruction here %
rec=InlineReconPipeline(twix_like);
%%%%%%%%%%%%%%%%%%%
Step 4: Modify your existing offline reconstruction script
Typical modifications:
• Raw data read-in
Demo:
% original code
data=mapVBVD(fileName);
% modified code
data=twix_like;
• [Optional] we also provide some useful tools for geometry computation and
(B)
Offline NUFFT reconstruction: executive main function
Inline implemented NUFFT reconstruction: main function
Wrap in a function to be inserted in the background recon handler
The pathes and input raw data has been handled by the framework, thus remove the related lines
The core function to call the toolbox Modifications shows in below figure
[Optional] The intermediate results could be saved in the same folder of raws
Offline NUFFT reconstruction: core recon function (partial)
Inline implemented NUFFT reconstruction: core recon function (partial)
Instead of reading by mapVBVD from Twix format raw, use the converted Twix-like raw

*Figure 2 A diagram to demonstrate the code-level modifications (in 4 steps) to translate an offline reconstruction for inline use. Essential modifications include: 1) modify the centralized configuration file (Framework_config.xml) to register a new pipeline (i.e., defining 'Recon_ID' and relative configurations); 2) define a Read&Save handler by copying the provided template and specifying the matching 'Recon_ID'; 3) define a background reconstruction handler similarly; 4) adapt the existing offline reconstruction script and wrap it as a function to insert in the background reconstruction handler. Since the ISMRMRD format raw has been converted back to a twix-like structure, the existing offline reconstruction script could be maximumly reused.*

Screenshots from the MR console demonstrate successful inline integration of AlignedSENSE on a T1-MPRAGE sequence. The native scanner-reconstructed image was returned for early review (Fig. 3A). The AlignedSENSE reconstruction (Fig 3B) proceeds on the remote computer whilst scanning continued uninterrupted, and once it was complete, the resulting images were retrieved via a retrieval scan (Fig. 3C). All sequences, including intermediate scans, were scanned and reconstructed without delay, confirming the framework's non-disruptive integration with standard clinical workflows.

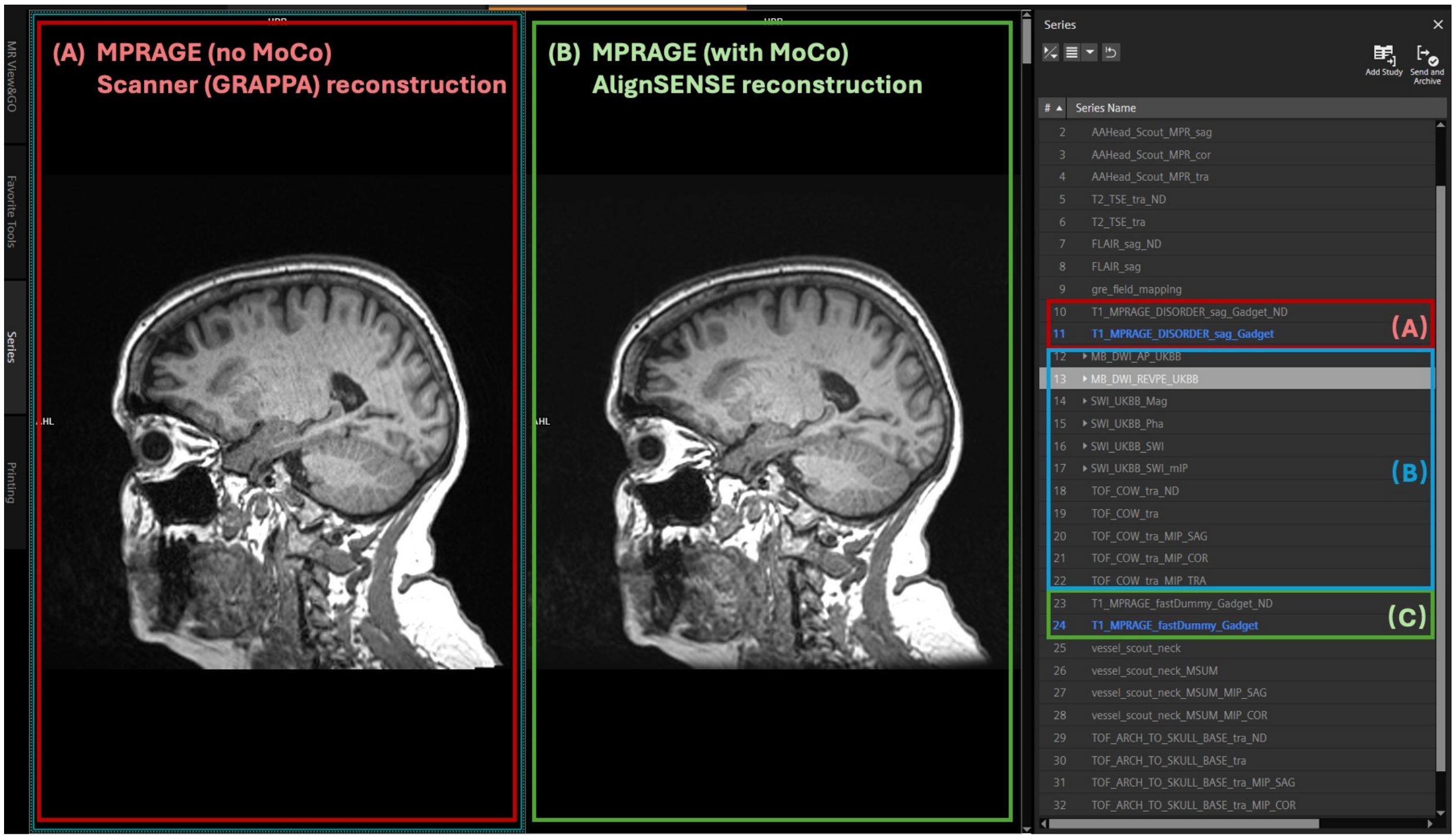

*Figure 3 Screenshots from the MR console illustrate a protocol in which the MPRAGE sequence was implemented with inline AlignedSENSE reconstruction via the proposed framework. The conventional scanner reconstruction (GRAPPA) was returned first for early review (A). Intermediate sequences proceeded without interruption in scanning or reconstruction (B). After the AlignedSENSE reconstruction completed, a retrieval scan was triggered to return the AlignedSENSE-reconstructed image to the MR console (C). Compared to the scanner-reconstructed image, the AlignedSENSE result demonstrated equivalent image properties, with the added benefit of effective removal of motion related artifacts.*

In Figure 4, compared with scanner-reconstructed image (left), the retrieved image (right) maintained consistency through the successfully applied scanner-based corrections (i.e., bias field and distortion correction for our case), which were absent in the non-retrieved image (middle) produced by the remote reconstruction code (note the image shading and lack of geometry correction in the neck, orange dashed arrow). Additionally, the framework's support for multiple inputs was demonstrated by eliminating a wrapping artifact by reconstructing to a larger FOV using an external reference scan for coil sensitivity estimation for SWI (Fig. 4B, yellow arrow).

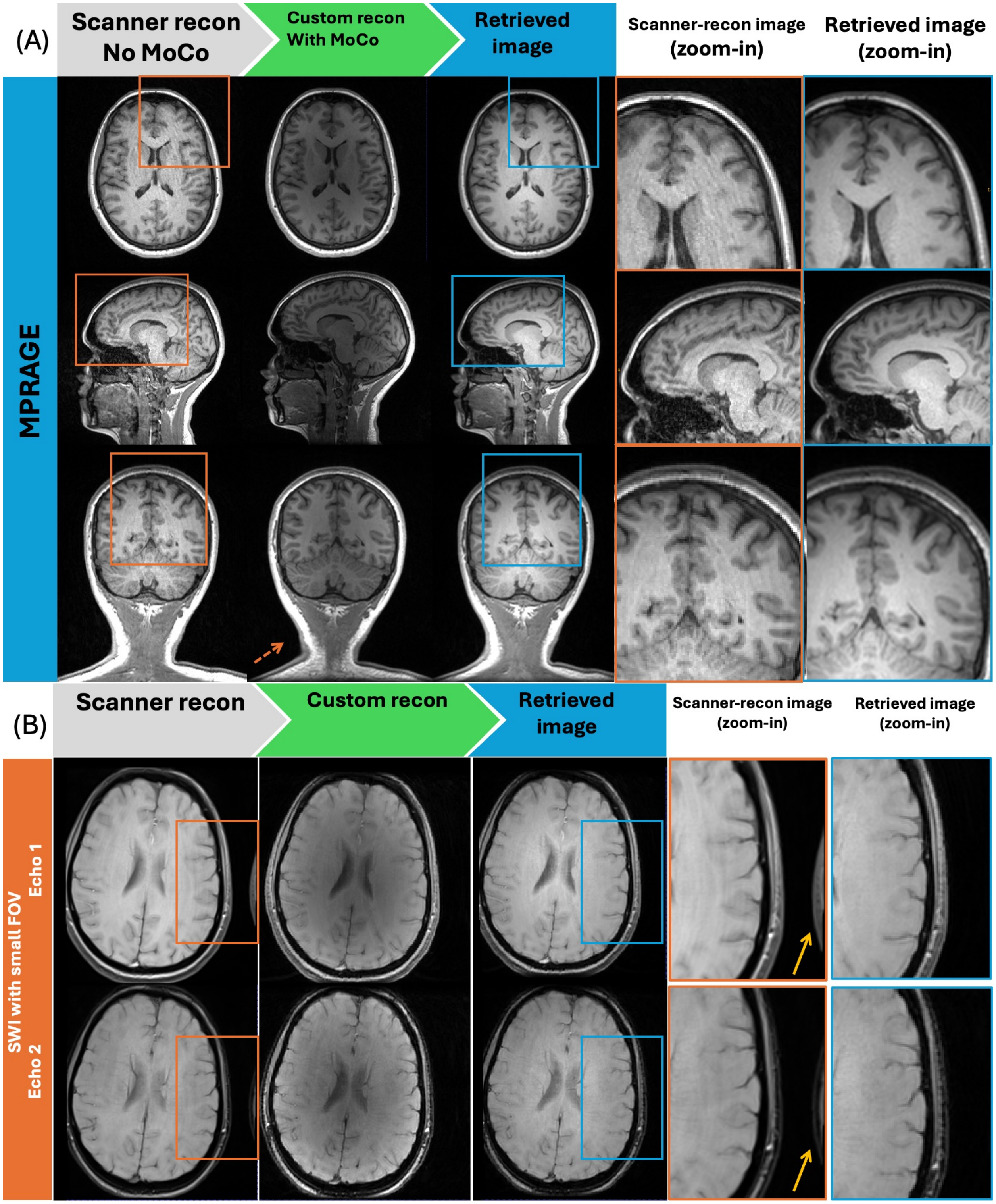

*Figure 4 Comparison of scanner-reconstructed, AlignedSENSE-reconstructed images before retrieval, and AlignedSENSE-reconstructed images after retrieval (from left to right) for the MPRAGE (A) and SWI (B) sequences. Successful application of scanner-based bias field and distortion correction can be observed on the retrieved images (orange dashed arrows), while the pre-retrieval AlignedSENSE images lack these standard post-processing steps, resulting in visible differences from the scanner reconstructions. Demonstrating the framework's capability to support multi-input reconstructions, the*

*wrapping artifact seen in the scanner-reconstructed SWI image (yellow solid arrow) was removed by using an external reference scan with a larger FOV for coil sensitivity estimation and reconstructing to a correspondingly larger FOV.*

## Feasibility for Multiple Inline-implemented Sequences within One Examination

The neuroimaging protocol with five sequences implemented inline via the proposed framework was successfully executed without disrupting scanning or reconstruction. From the timeline (Fig. 5A), all sequences ran sequentially without delay. Standard scanner reconstructions provided immediate console review (Fig. 5B, top row). Motion-corrected reconstructions (“Moco recon”) for the first three sequences were distributed across three GPUs and processed in parallel. The final sequence was queued and automatically launched once resources became available, demonstrating the framework’s resource-aware scheduling. Finally, all custom-reconstructed images were successfully retrieved using retrieval scans (Fig. 5B, button row).

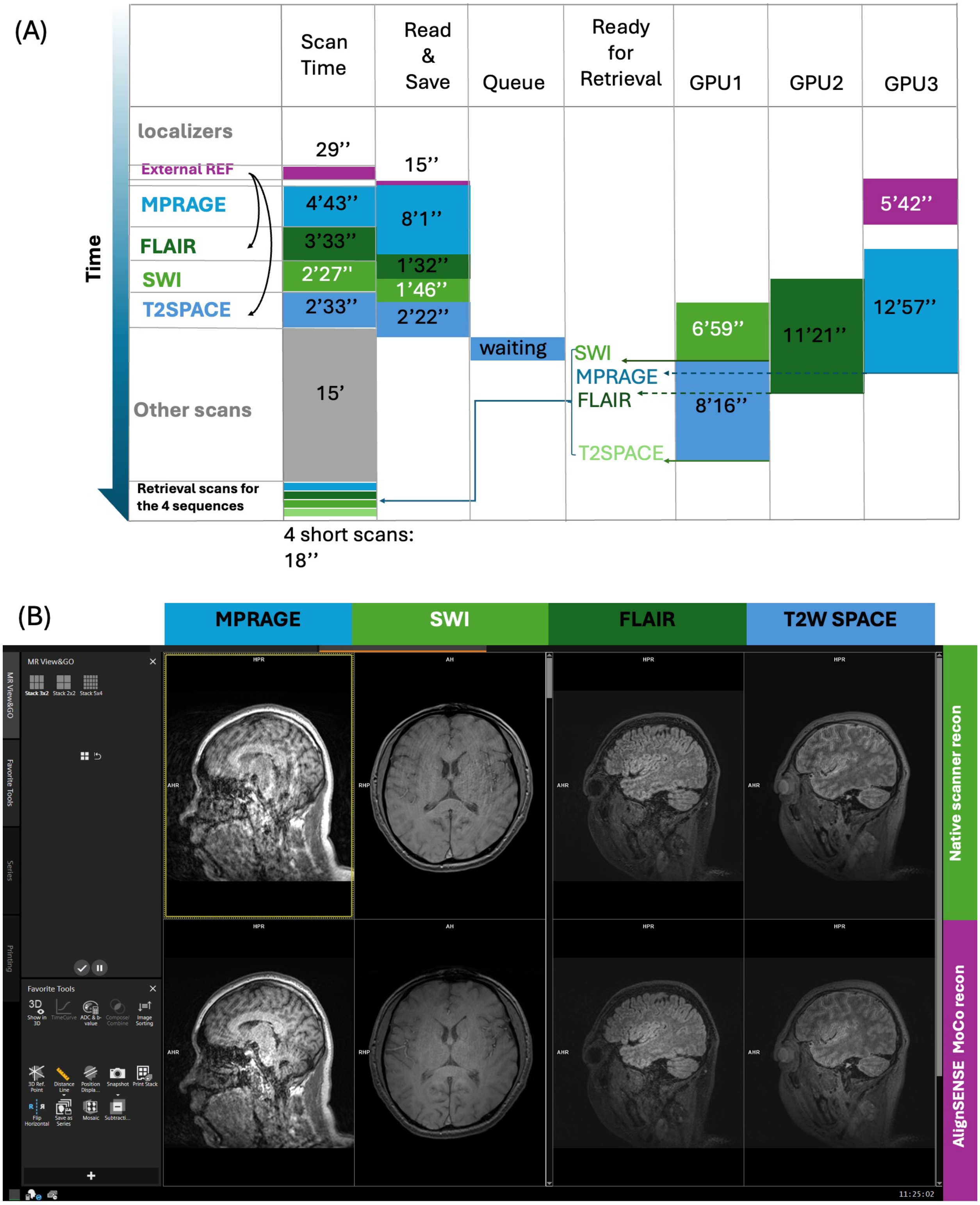

*Figure 5 Example of resource management and ability asynchronous reconstruction framework to enable uninterrupted scanner operation while the remote server completes its tasks. (A) Timeline diagram for the full protocol showing the four sequences implemented inline AlignedSENSE reconstruction for motion correction and the external reference for coil*

*sensitivity estimation, and (B) screenshots on the MR console with scanner-reconstructed images of the four sequences in the top row and inline AlignedSENSE-reconstructed images in the bottom row. Data from the first three sequences was reconstructed in parallel on three GPUs, while data from the fourth sequence was automatically queued. After the first sequence completed, the fourth sequence reconstruction was triggered in parallel. Finally, the reconstructed images saved on the server were retrieved to the MR console via retrieval scans.*

## Robustness

Between 3 April 2024 and 4 April 2025, 480 subjects were scanned for the TwinsUK Project using a neuroimaging protocol that included T1-MPRAGE with inline motion correction (AlignedSENSE) reconstruction. No disruptions or delays were reported. A retrospective server database review showed that motion-corrected images were successfully generated for all 480 cases. A retrospective PACS review showed that 476 cases (99%) included both scanner and motion-corrected images. Of these, 456 cases (95%) were retrieved during the examination via a retrieval scan, and 21 cases (4%) were retrieved using retro-reconstruction afterward. In four cases (<1%), only scanner-reconstructed images were available in PACS.

For those four cases, reconstruction on the external server was completed, but the retrieval scans were triggered before images were ready due to extended reconstruction times (e.g., severe motion). Although retro-reconstruction could have recovered the images, this was not done at the time and the original raw data was not retained, so this pathway closed. Since images were reconstructed on the remote server, use of a retrieval scan remained feasible, but without scanner-based post-processing. These cases highlight the need to choose between retrieval scans and retro-reconstruction according to the characteristics of each method, which are discussed in detail in Supplementary Material A.

# Discussion

In this work, we have presented an open-source framework for inline deployment of established offline MR reconstructions using Gadgetron. The framework simplifies integration by addressing key practical barriers through targeted design features: data format incompatibility is resolved enabling maximum script reuse; multi-input and long reconstruction are supported by enabling an asynchronous trigger-and-retrieval mechanism, resource-aware scheduling, and integrated data management. We validated feasibility and robustness through representative reconstructions and a large-scale clinical study, with all components and demonstration examples openly released to support adoption and further development.

While the framework addresses key barriers to inline integration, several limitations remain, so it is still under active development.

First, the framework currently depends on a full Gadgetron installation. While it has the potential to be extended to Gadgetron-style platforms such as FIRE™ and OpenRecon™, its current implementation is specific to Gadgetron. To ease setup, detailed instructions for Gadgetron environment deployment are provided in the user manual[14]. For platforms that lack core Gadgetron features, the framework's logic may still offer reference value, although the provided scripts might not be directly applicable.

Second, the input converter generates a Twix-like structure, limiting the framework's use to Siemens platforms, although other components are vendor-independent. Additionally, the current Twix-like data structure covers most common cases, but the ParameterMap does not yet capture all headers from the native Twix format, which may affect scripts relying on less common header fields. For example, geometry-related metadata have not yet been fully mapped, and dedicated utility functions are currently provided to generate alternative parameters. This gap is expected to be closed in future updates. Clear instructions are provided for extending the ParameterMap within the manual[14], and community contributions are invited to improve generalizability of the framework.

Third, the framework is currently implemented in MATLAB, so that reconstruction programs developed in other programming languages or as standalone executables need to be embedded within MATLAB scripts for compatibility, which is not always straightforward. A Python-based version is in development to broaden accessibility.

Fourth, resource monitoring currently focuses on GPU availability. While sufficient for GPU-intensive reconstructions, the same logic can be extended to monitor CPU and RAM usage.

Looking ahead, as interest in open platforms and inline reconstruction grows, issues such as data format incompatibilities may recede if standardized interchange specifications, such as ISMRMRD, become the norm for both offline and inline deployment. It remains to be seen whether other frameworks will offer the same flexibility for multi-input and asynchronous reconstruction with simple retrieval pathways to the scanner console and database, enabling inline deployment without affecting examination structure or operation.

In summary, the proposed framework lowers the technical barrier for inline deployment of established offline reconstructions with minimum modification, regardless of input numbers or reconstruction time. It was validated as suitable for both routine clinical use and large-scale cohort studies.

## Acknowledgments

We acknowledge support from the Gadgetron development team for detailed consultations on technical aspects of Gadgetron.

We acknowledge support of all colleagues contributed to the TwinsUK Projects and its fundings: the Wellcome Trust, Medical Research Council, Versus Arthritis, European Union Horizon 2020, Chronic Disease Research Foundation (CDRF), Wellcome Leap Dynamic Resilience Programme (co-funded by Temasek Trust),  Zoe Ltd, the National Institute for Health and Care Research (NIHR) Clinical Research Network (CRN) and Biomedical Research Centre based at Guy's and St Thomas' NHS Foundation Trust in partnership with King's College London.

LCG acknowledges funding from Project PID2024-162095OB-I00, funded by MICIU/AEI/10.13039/501100011033 and co-financed by FEDER, UE.

# References

1. Yarra Framework. 2025. Accessed June 25, 2025. https://yarra-framework.org/
2. Hansen MS, Sørensen TS. Gadgetron: An open source framework for medical image reconstruction. *Magn Reson Med*. 2013;69(6):1768-1776. doi:10.1002/mrm.24389
3. Gadgetron GitHub. Accessed June 25, 2025. https://github.com/gadgetron
4. O. Afacan. Online Open Source MRI Reconstruction Platforms. In: *ISMRM & ISMRT Annual Meeting and Exhibition* . ; 2025.
5. Chow K, Kellman P, Xue H. Prototyping image reconstruction and analysis with FIRE. In: *In Proceedings SCMR 24th Annual Scientific Sessions*. ; 2021.
6. OpenRecon | Exceleration Edition - Siemens Healthineers. 2025. https://www.siemens-healthineers.com/en-uk/magnetic-resonance-imaging/options-and-upgrades/upgrades/excelerate-edition-upgrade
7. GyroTools | ReconFrame. 2023. Accessed June 25, 2025. https://www.gyrotools.com/gt/index.php/products/reconframe
8. Orchestra | Python SDK. 2025. Accessed June 25, 2025. https://docs.getorchestra.io/docs/integrations/utility/http/python-sdk
9. Inati SJ, Naegele JD, Zwart NR, et al. ISMRM Raw data format: A proposed standard for MRI raw datasets. *Magn Reson Med*. 2017;77(1). doi:10.1002/mrm.26089
10. Siemens to ismrmrd | GitHub. Accessed August 7, 2025. https://github.com/ismrmrd/siemens_to_ismrmrd
11. GE to ismrmrd | GitHub. Accessed August 7, 2025. https://github.com/ismrmrd/ge_to_ismrmrd
12. Philips to ismrmrd | Zenodo. Accessed August 7, 2025. https://zenodo.org/records/32595
13. Uus AU, Neves Silva S, Aviles Verdera J, et al. Scanner-based real-time 3D brain+body slice-to-volume reconstruction for T2-weighted 0.55T low field fetal MRI. Published online April 23, 2024. doi:10.1101/2024.04.22.24306177
14. Practical_Inline_Recon_Framework-public | GitHub. Accessed August 7, 2025. https://github.com/ZihanNing/Practical_Inline_Recon_Framework-public.git
15. Pruessmann KP, Weiger M, Scheidegger MB, Boesiger P. SENSE: sensitivity encoding for fast MRI. *Magn Reson Med*. 1999;42(5):952-962.
16. Cordero-Grande L, Teixeira RPAG, Hughes EJ, Hutter J, Price AN, Hajnal J V. Sensitivity Encoding for Aligned Multishot Magnetic Resonance Reconstruction. *IEEE Trans Comput Imaging*. 2016;2(3). doi:10.1109/TCI.2016.2557069
17. Cordero-Grande L, Ferrazzi G, Teixeira RPAG, O'Muircheartaigh J, Price AN, Hajnal J V. Motion-corrected MRI with DISORDER: Distributed and incoherent sample orders for reconstruction deblurring using encoding redundancy. *Magn Reson Med*. 2020;84(2). doi:10.1002/mrm.28157

18. Fessler JA, Sutton BP. Nonuniform fast Fourier transforms using min-max interpolation. *IEEE Transactions on Signal Processing*. 2003;51(2). doi:10.1109/TSP.2002.807005
19. Blunck Y, Josan S, Taqdees SW, et al. 3D-multi-echo radial imaging of 23 Na (3D-MERINA) for time-efficient multi-parameter tissue compartment mapping. *Magn Reson Med*. 2018;79(4):1950-1961. doi: 10.1002/mrm.26848.
20. Rot, S, Cleary J, Dokumaci AS, et al. (2023). Whole-brain bi-exponential 23Na-MRI T2* mapping at 7T with a 32-channel phased array receiver coil. Proceedings of the International Society for Magnetic Resonance in Medicine (ISMRM), Abstract 0021.
21. Uecker M, Lai P, Murphy MJ, et al. ESPIRiT - An eigenvalue approach to autocalibrating parallel MRI: Where SENSE meets GRAPPA. *Magn Reson Med*. 2014;71(3). doi:10.1002/mrm.24751
22. Verdi S, Abbasian G, Bowyer RCE, et al. TwinsUK: The UK Adult Twin Registry Update. *Twin Research and Human Genetics*. 2019;22(6). doi:10.1017/thg.2019.65
23. TwinsUK – The biggest twin registry in the UK for the study of ageing related diseases. https://twinsuk.ac.uk/.

# Supplementary Materials

## Section A: Retrieval Scans vs Retro-reconstruction

Two strategies are available for image retrieval: a retrieval (short dummy) scan or retro-reconstruction.

Retrieval scans acquire k-space with the same matrix size of the custom-reconstructed images to enable console reception. Since the acquired data are discarded afterwards, the dummy sequence can be highly accelerated; or in our case, it was shortened to only acquire a single central k-space line (< 5 seconds). To note, some DICOM headers would be generated based on the dummy scan instead of the target scan. This approach allows fully automated integration into the scanning protocol by placing the dummy sequence at the end of the protocol or at a timepoint when reconstruction is expected to finish. However, since that the dummy sequence is eventually not sending out the same data as the data for custom reconstruction, three conditions must be met: (1) (in default settings) the retrieval scan's name should contain the target scan's name for correct allocation of the image; (2) the matrix size and related parameters (e.g., oversampling and partial Fourier factors) should match the custom image; (3) the scan position should match the target scan, as some scanner post-processing (e.g., bias field correction) requires matched dependent lines (e.g., noise and calibration scans).

Retro-reconstruction, in contrast, inherently preserves all relevant settings (e.g., sequence name, matrix size, dependent lines) and thus avoids these constraints. Its limitation is the need for manual pipeline selection on the MR console interface, which reduces automation.

Experience from the TwinsUK project illustrates their complementary roles. Retrieval scans are well suited for routine use as they enable fully automated image return without manual configuration. However, scanner-based post-processing can fail if the planning position does not match the target scan, which means it is no longer feasible once the subject has left the table. In contrast, retro-reconstruction provides greater flexibility but requires manual configuration, so it is used mainly as a fallback when retrieval scans fail (e.g., no image is ready when retrieval scan triggered). Yet, if raw data are not saved, as in the TwinsUK project, retro-reconstruction ceases to be available, emphasizing the importance of coordinated timing and data retention for robust clinical deployment.

Notably, if a retrieval scan is triggered before custom reconstruction finishes, no images are retrieved, although an error log is generated. Neither scanning nor other reconstructions are disrupted, and retrieval scans or retro-reconstructions can be repeated. Additionally, since retrieval scans and retro-reconstructions are repeatable, multiple reconstructions for the same scan can be run in parallel on the server without reacquiring data, enabling potential user scenarios.

Table S1. The inline deployment details of SENSE, AlignedSENSE, and NUFFT reconstructions

| Reconstructions deployed inline | MR scanner and sequences | External server |
|---|---|---|
| SENSE [12] and AlignedSENSE [13,14] reconstructions | 3T MR scanner (MAGNETOM Vida, Siemens Healthineers, Forchheim, Germany) running XA60 software, using SWI and MPRAGE sequences | Intel® Xeon® Silver 4314 CPU<br>125 GB RAM, 32 GB swap memory<br>Three NVIDIA GeForce RTX A6000 GPUs (48 GB memory each) |
| NUFFT reconstruction[18] for $^{23}Na$ imaging[19,20] | 7T MR scanner (MAGNETOM Terra.X, Siemens Healthineers, Forchheim, Germany) running XA60 software with a multi-echo GRE sequence with radial trajectory | Intel® Core™ i9-10920X CPU<br>125 GB RAM, 64 GB swap memory<br>Two NVIDIA GeForce RTX 2080 Ti GPUs (11 GB memory each) |

Table S2. The sequence parameters and reconstruction information of the demonstration sequences for inline implementation

| | External reference | SWI | T1-MPRAGE | Radial multi-echo GRE |
|---|---|---|---|---|
| Field strength | 3 T | 3 T | 3 T | 7 T |
| Field of view, $mm^3$ | 256×256×288 | 230×204×144 | 256×256×208 | 200×200×200 |
| Resolution, $mm^3$ | 4×4×4 | 0.8×0.8×3 | 1×1×1 | 3.1×3.1×3.1 |
| TE(/ΔTE)/TR, ms | 1.55/417 | 9.42/10.28/27 (2 echoes) | 1.97/2000 | 0.35/4/151 (38 echoes) |
| TI, ms | N.A. | N.A. | 880 | N.A. |
| Flip angle, deg | 4 | 15 | 8 | 90 |
| Acceleration | None | GRAPPA 2×1 | GRAPPA 2×1 | None |
| Partial Fourier | N.A. | Phase: 7/8<br>Slice: 7/8 | N.A. | N.A. |
| Turbo factor | 72 | N.A. | 208 | N.A. |
| DISORDER related[14] | N.A. | N.A. | Tile size, 15×9 Random-checkered | N.A. |
| Scan time | 29s | 2min 59s | 4min 43s | 15min 06s |
| Custom reconstruction | ESPIRiT for coil sensitivity estimation[16] | SENSE reconstruction[12] | AlignedSENSE reconstruction[13,14] | NUFFT[18,19,20] |
| Reconstruction time* | ~6min | ~12min | ~13min | ~3min |

* To note, the reconstruction time depends on the size and complexity of the raw data (e.g., motion severity) as well as the available computational resources.

Table S3. The sequence parameters and reconstruction information of the protocol

| | SWI (smaller FOV) | FLAIR | T2 SPACE |
|---|---|---|---|
| Field of view, $mm^3$ | 180×160×144 | 250×250×160 | 250×250×160 |
| Resolution, $mm^3$ | 0.8×0.8×3 | 1×1×1 | 1×1×1 |
| Imaging view | Transverse | Sagittal | Sagittal |
| TE(/ΔTE)/TR, ms | 9.42/10.28/27 (2 echoes) | 394/8000 | 408/3200 |
| TI, ms | N.A. | 2300 | N.A. |
| T2 Prep duration, ms | N.A. | 125 | N.A. |
| Flip angle, deg | 15 | T2 Variable Scheme | T2 Variable Scheme |
| Acceleration | GRAPPA 2×1 | CAIPIRIHNA 2×2 | CAIPIRIHNA 2×2 |
| Partial Fourier | Phase: 7/8 Slice: 7/8 | N.A. | N.A. |
| Turbo factor | N.A. | 278 | 278 |
| DISORDER related[14] | Tile size, 5×8 Random-checkered | Tile size, 8×3 Random-checkered | Tile size, 8×3 Random-checkered |
| Scan time | 2min 8s | 3min 33s | 1min 33s |
| Custom reconstruction | AlignedSENSE reconstruction[13,14] | | |
| Reconstruction time† | ~7min | ~12min | ~8min |

* The sequence parameters and reconstruction information of the external reference and T1-MPRAGE sequences were the same as shown in Table S2

† The reconstruction time depends on the size and complexity of the raw data (e.g., motion severity) as well as the available computational resources.